# Understanding Developers Privacy Concerns Through Reddit Thread Analysis

Jonathan Parsons[1,*,†], Michael Schrider[1,†], Oyebanjo Ogunlela[1] and Sepideh Ghanavati[1]

[1]*The University of Maine, Orono, ME 04469*

### Abstract

With the growing global emphasis on regulating the protection of personal information and increasing user expectation of the same, developing with privacy in mind is becoming ever more important. In this paper, we study the concerns, questions, and solutions developers discuss on Reddit forums to enhance our understanding of their perceptions and challenges while developing applications in the current privacy-focused world. We perform various forms of Natural Language Processing (NLP) on 437,317 threads from subreddits such as r/webdev, r/androiddev, and r/iOSProgramming to identify both common points of discussion and how these points change over time as new regulations are passed around the globe. Our results show that there are common trends in privacy topics among the different subreddits while the frequency of those topics differs between web and mobile applications.

## 1. Introduction

As the world is continuously advancing in ever-growing connected ways, software developers and requirements analysts are required to implement privacy-preserving solutions to protect users' privacy in their applications. On the other hand, users are becoming more conscious about how their information is collected and used by various organizations. These parallel growths resulted in the creation of several privacy-focused regulations, such as the General Data Protection Regulation (GDPR) [1] and California Consumer Protection Act (CCPA) [2].

To ensure compliance with these regulations, systematic privacy by design approaches and tools need to become the norm. Without proper privacy education, tools, and guidelines, developers look into forums such as Stack Overflow [3], or Reddit [4] to find solutions to their privacy-related questions. Understanding the types of privacy-focused questions asked on these forums and developers' challenges helps better tailor the tools and approaches to their needs.

*Corresponding author.

†These authors contributed equally.

Envelope-Open jonathan.parsons1@maine.edu (J. Parsons); michael.schrider@maine.edu (M. Schrider); oyebanjo.ogunlela@maine.edu (O. Ogunlela); sepideh.ghanavati@maine.edu (S. Ghanavati)

GLOBE https://www.sepidehghanavati.com/ (S. Ghanavati)

Orcid 0000-0003-2778-8136 (J. Parsons); 0000-0002-3725-7438 (M. Schrider); 0000-0000-0000-0000 (O. Ogunlela); 0000-0001-7972-667X (S. Ghanavati)

In addition to surveys [5], in recent years, some work focused on evaluating privacy-related questions on Stack Overflow (SO) and Reddit [6, 7, 8]. Li et al. [8] conducted an analysis of Android developers' privacy concerns on r/anddroidev subreddit and discovered that privacy appears to be underrepresented. Tahaei et al. [6, 7] evaluated SO and show that conversations are mostly focused on compliance with regulations, often citing official documents. Li et al. [9] used Reddit [10] to identify the narratives driving users' privacy concerns rather than developers and showed that understanding the users' concerns could drive developers' concerns, as well. In this paper, we will extend Li et al. [8] approach by evaluating other similar subreddits, /r/iOSprogramming and /r/webdev, and conduct a comparative analysis on the types of privacy questions asked based on privacy requirements imposed by these three frameworks (i.e., Android, iOS, and Web) and the role of privacy regulations on the questions. We also examine the sentiment regarding privacy, comparing that amongst various subreddit communities to get a better idea about how developers actually view privacy, not just how they comply with it. Our analysis of 437,317 threads shows that there are differences in the most frequent topics based on mobile and web app development. We also observe that the regulations such as GDPR and CCPA impacted the topics' trend, however, we could not conclude a change in the overall sentiment. Our research questions which we will answer in this paper are as follows:

- **RQ1:** What are the major privacy concerns on developer forums?
- **RQ2:** What is the overall sentiment of developing privacy requirements in the studied communities? How does it differ by subreddits?
- **RQ3:** How do regulations such as GDPR and CCPA influence RQ1 and RQ2?

## 2. Related Works

In recent years, several works focused on evaluating and understanding the privacy challenges of developers and their approach to implementing privacy requirements. Some studies evaluate developers' privacy behaviors through questionnaires or surveys [11, 12, 13] and show that developers generally see privacy as a burden and afterthought and are not familiar with basic privacy concepts [14, 5]. Other research [6, 7, 8, 15] evaluate developers' popular forums such as Stack Overflow (SO) [3] and Reddit [4] to identify the topics of developers' questions.

Proferes et al. [10] analyzed 727 publications between 2010 and 2020 and identified a substantial increase in the usage of Reddit as a data collection medium by disciplines ranging from computer science to social sciences. Iqbal et al. [16] evaluated the data from the top 10 mobile and desktop apps' subreddits and identified that 54% of the post included useful information such as bug reports or feature requests and could be used for requirements elicitation.

Analysis of 4,957 Reddit comments in 180 security- and privacy-related discussions from /r/homeautomation show that users' concerns are context-dependent and their attitude towards privacy and security can change during the different phases of adoption of smart home devices [17]. Li et al. [8] analyzed 207 discussions on r/androiddev subreddit to identify how developers discuss personal data protection. Their findings indicate that privacy concerns are not discussed often on developer's forums and developers shy away from discussions relating to privacy concerning plan and execution trials. However, they posit that they seem externally motivated by new demands for privacy emerging from privacy-focused regulations. Tahaei et al. [6, 7, 18] studied 315 privacy questions on SO and identified that the introduction of Google and Apple privacy labels resulted in an increase in the number of privacy-related questions on SO [6]. In another study [7], they explored the types of advice given by developers on privacy issues and compared them with Hoepman's approach [19]. They identified 148 pieces of advice focused on regulations and confidentiality, including 'inform,' 'hide,' 'control,' and 'minimize'.

In this paper, we extend the current approaches [7, 8, 11, 18] in two ways: (a) the number of subreddits to review and (b) the scope at which they are analyzed. We extend Li et al. [8] by adding r/iOSprogramming and r/webdev to r/androiddev to get a more general idea of developers' privacy concerns. These subreddits constitute the majority of the current software development. We also conduct *sentiment* analysis on the discussions similar to [20, 21, 22], and evaluate how those may change between the different communities to gain insight into the emotions behind the developers when it comes to privacy and the surrounding policy.

# 3. Methodologies

In this section, we describe our methodology for collecting and creating the Reddit dataset (i.e., Section 3.1) and then explain the process of analyzing the data (i.e., Section 3.2).

## 3.1. Gathering Reddit Data

To gather data, we utilize Pushshift Multithread API Wrapper [23, 24] which allows for granular information and provides the ability to multithread, in compare to other work [9, 20, 25].

The current related work provides insights regarding *what* and *how* of extracting the information. Li et al. [8] focus on /r/androiddev, which includes 203k members (top 1% of subreddits by size) [26] and is an active community with the discussions limited to high-level android app design. We extend their effort to two other similar subreddits, /r/iOSprogramming (120k members) [27] and /r/webdev (1.4mil members) [28] to broaden the scope of the analysis of developers' privacy concerns across various development platforms. In total, we pulled 100,040 submissions from r/androiddev, 55,553 submissions from r/iOSprogramming, and 281,724 submissions from r/webdev along with all associated comments, from January 2014 through November 2022. We chose these subreddits specifically due to their relative similarity in being platform-specific, developer-centric forums.

## 3.2. Processing and Analyzing Reddit Data

After collecting the data, we need to identify privacy-related submissions. A common way to do this is by filtering posts and discussions that contain the term "privacy" [6, 7], or other similar terms that are often used in discussions around privacy such as "GDPR", "CCPA", "mac address", "location", etc [8, 9, 21]. In Table 1, we augmented the terms identified by Li et al. [8] (i.e., Unique Identifier, Photo, and Video, Audio, and Location) with general privacy and privacy regulation terms (i.e., Privacy category in Table 1). We then used these keywords to create a privacy-related dataset before performing our analysis. We also preprocessed the data by removing stopwords, stemming, and lemmatizing text to prepare for our analysis [29].

**Table 1**
A List of the Privacy Terms

| Data Category | Privacy Keywords |
|---|---|
| Privacy | private, privacy, gdpr, general data protection regulation, ccpa, ccpr, califronia consumer privacy act, California consumer privacy regulation |
| Unique Identifier | first name, last name, real name, identification number, id number, social security number, ssn, license number, passport number, screen name, user name, account name, user id, username, userid, online identifier, imei, 1 device serial number, advertising id, android id, ssaid, mac address, imsi, instance id, guid, internet protocol address, ip address, email address, telephone number, phone number, line1 |
| Photo and Video | video recording, camera, gallery |
| Audio | audio recording, microphone, voice |
| Location | voice, location, physical street address, home address, physical address, street name, city name, postal address |

To answer RQ1, we conducted simple phrase frequency analysis and then identified the topics of each submission by leveraging Latent Dirichlet Allocation (LDA) [30] similar to other work [29]. We first narrowed down our dataset to privacy questions via an Adaptive Boosting (AdaBoost) model to classify posts in our dataset as questions. The AdaBoost classifier contained 25 decision stumps and was trained on a combination of subsets of SQuAD and SPAADIA datasets which include phrases and sentences labeled as either statements or questions [31, 32, 33]. We transformed the training data into matrices of token counts and validated the classifier against 200 randomly sampled posts. We manually classified posts from AndroidDev containing 93 (46.5%) questions and 107 (53.5%) statements. The classification was performed by a single team member and verified by another; the agreed-upon method was to read both the title and body, and if either had a question from the redditer to the community, then it was classified as a question; framing and rhetorical questions were not considered questions. Against the validation set, our classification achieved 71% accuracy. Through LDA analysis, we generated ten topics of four words each for posts Pre/Post GDPR and Pre/Post CCPA.

To address RQ2, we use qualitative metrics to evaluate both the discussions and posts [10]. Qualitatively, we perform sentiment analysis leveraging Natural Language Toolkit (NLTK) approaches [20, 21, 34].

We answer RQ3 by applying the same term frequency analysis and LDA topic analysis used for RQ1 against datasets filtered to pre- and post-regulations and trending the RQ2 sentiment analysis to examine if there is any change due to the introduction of GDPR (April 2016) and CCPA (June 2018).[2]

## 4. Results

In the first step, we conducted a simple word counting analysis to identify how often privacy-related words appear in the initial posts of r/androiddev, r/iOSProgramming, and r/webdev forums. We noticed that typical permission requests, such as “location”, “camera”, “gallery”, “microphone” and “voice” are the more prevalent words, though they still make up a small portion of the overall posts. Additionally, we observed that the most common word, *location*, appears nearly twice as often as the second common one, *camera*. With more detailed analysis, we noticed that while *location* and *camera* are privacy-related words, they are more often referred to for non-privacy reasons; such as the best methods for having an app interact with a camera. After this high-level evaluation, we delve into our research questions.

[2]The data, models, and analysis are: https://github.com/mschrider/PEP_Privacy_Dev_Forum_Analysis/.

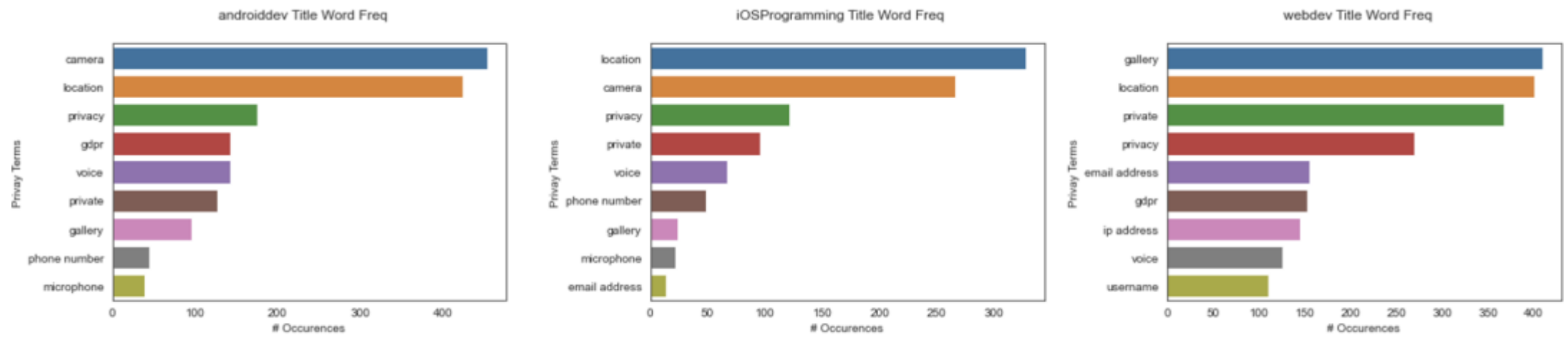

(a) Android Title Word Frequency (b) iOS Title Word Frequency (c) Web Title Word Frequency

**Figure 1:** Comparison of Title Text Frequency

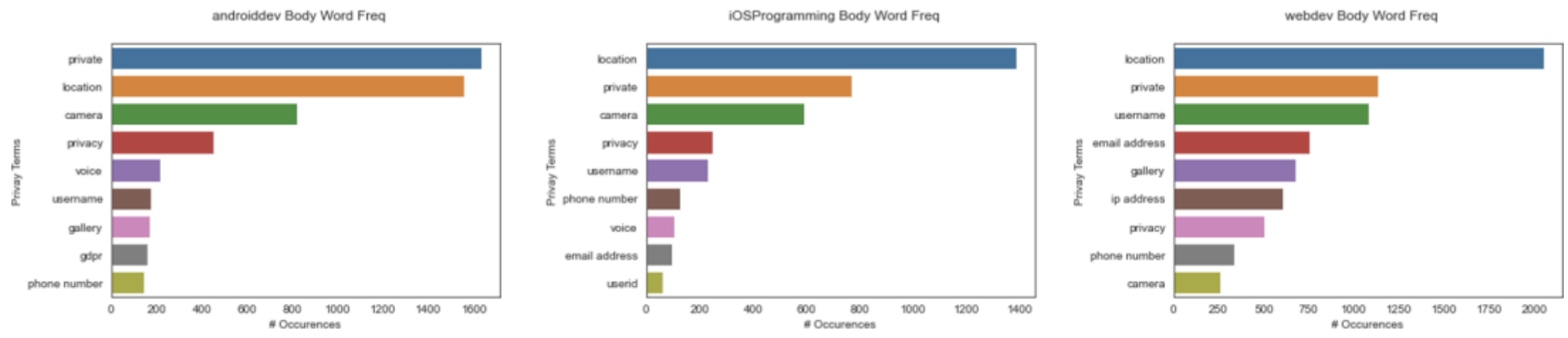

(a) Android Body Word Frequency (b) iOS Body Word Frequency (c) Web Body Word Frequency

**Figure 2:** Comparison of Body Text Frequency

## RQ1 - What are the major privacy concerns on developer forums?

The initial phrase frequency analysis indicates that “location” and “camera” are the main privacy-related terms referenced in the titles and bodies of posts. Meanwhile /r/webdev prioritizes “location” but replaces “camera” with “gallery” while “email address” and “username” are much higher up in the list (See Fig. 1 and Fig. 2). This data shows the inherent difference between the primary functions of web-based and mobile-based applications.

Based on the trends above, we narrowed down the questions to those containing “location” or “camera” for further analysis. We observe that most questions are not in fact privacy-related; though for *location*, one of the most upvoted questions is “GDPR - What all do I need to do?”. In general, we identified that the questions that were actually privacy-related were centered around asking for and granting permission to apps for capabilities that have impacts on privacy. Our LDA analysis shows common topics among privacy questions. Fig. 3 and Fig. 4 include the top four webdev topics for pre- and post-GDPR/CCPA. We observed that 'gdpr' is one of the top trending topics from its expected non-existence pre-GDPR. Topics with variations of the word policy, consent, or cookie increased in frequency post-GDPR/CCPA but privacy-related topics appear more frequently post-CCPA. There is an overlap in post-CCPA with post-GDPR data; thus, it may be hard to distinguish the effects of each regulation in isolation.

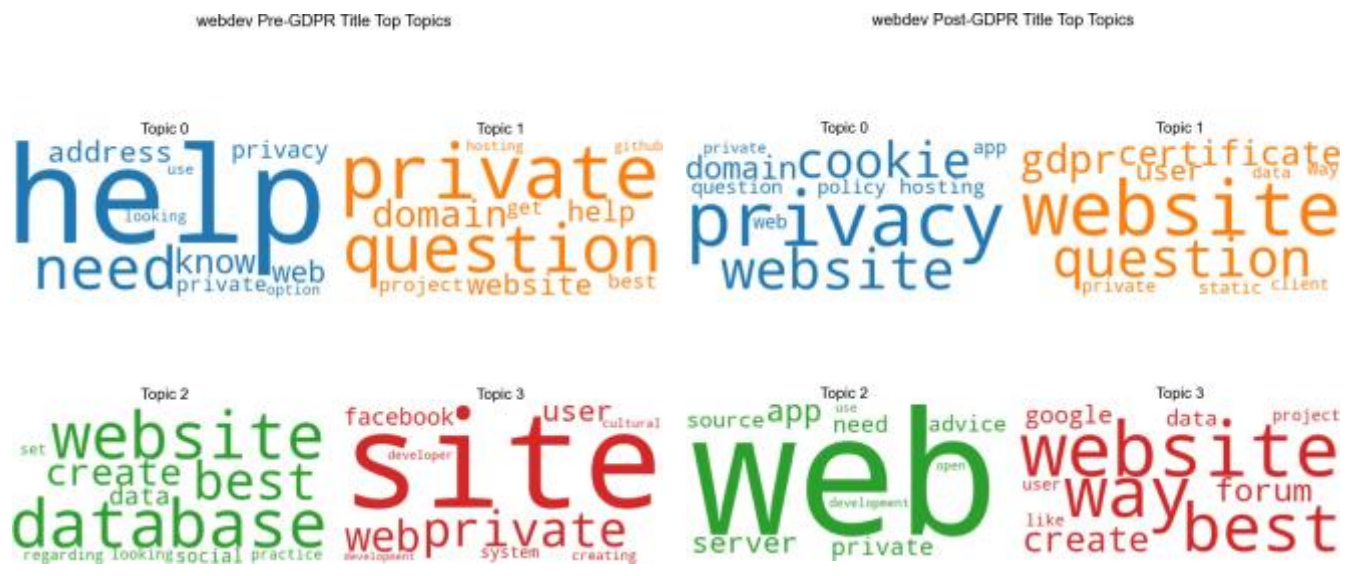

**Figure 3:** webdev Top Topics Pre (left) and Post (right) GDPR.

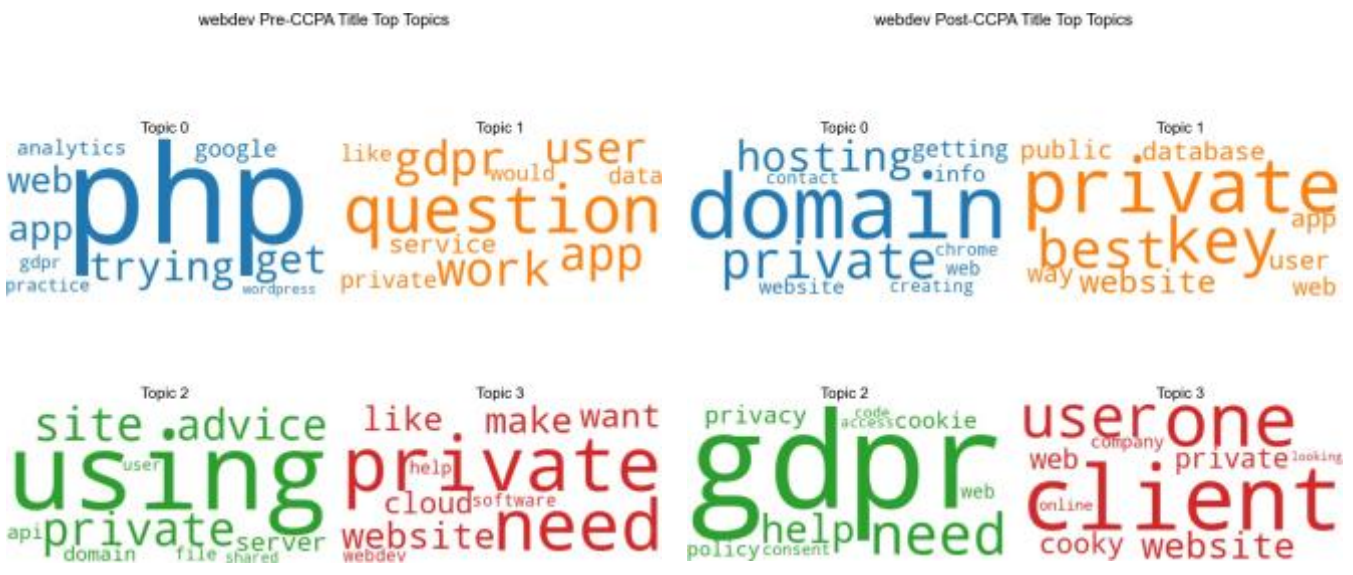

**Figure 4:** webdev Top Topics Pre (left) and Post (right) CCPA.

## RQ2 - What is the overall sentiment of developing privacy in the studied communities? Does it differ by subreddit?

Overall, all subreddits show similar sentiment profiles, with titles generally tending toward neutral and the main bodies of text tending toward positive. The results in Fig. 5 show the trends over time, with markers for where GDPR and CCPA were introduced. While all subreddits show undulations, there is no concrete evidence of trends in sentiment shifting one way or another.

## RQ3 - How do regulations such as GDPR and CCPA influence RQ1 and RQ2?

Topics and terms show a significant change due to GDPR but to a lesser extent CCPA. As shown in Fig. 3 and Fig. 4, the overall distribution of topics/terms changed noticeably after both regulations were enacted. However, despite the changing topics, we did not observe conclusive evidence of a change in sentiment post-GDPR/CCPA. All three subreddits showed very steady sentiment, with roughly similar noise values, across the entire timeline represented in the data.

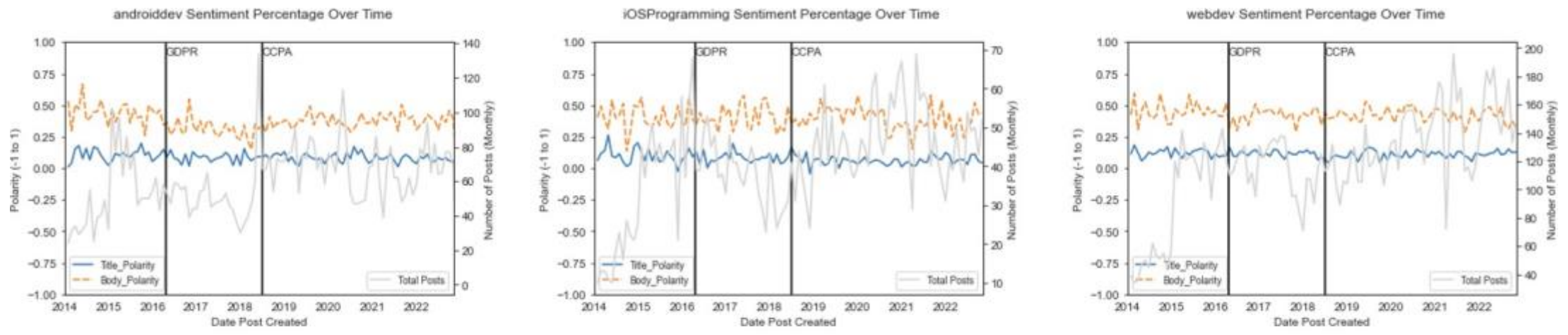

(a) Android Sentiment Over Time

(b) iOS Sentiment Over Time

(c) Web Sentiment Over Time

**Figure 5:** Sentiment over time for androiddev, iOSProgramming, and webdev subreddits.

## 5. Discussion

Similarities between /r/iOSProgramming and /r/androiddev are expected since both deal with mobile development. Mobile privacy concerns such as camera, microphone, and location permissions dominate the discussions while on /r/webdev, main focus is on cookies and websites. Understanding the trends and commonalities regarding developers' privacy concerns is important to platform owners (e.g., Apple, Google, etc.), regulators, and requirements analysts. Platform owners should provide support to mitigate and resolve developers' privacy questions. Regulations drive developers' concerns; therefore, regulators should study the impact of their regulations on developer behaviors. With this information, requirements analysts could focus on developing approaches to elicit and model privacy requirements from regulations and automate the compliance process.

This study was done solely on Reddit, which may introduce an inherent selection bias; for example, due to focusing on issues brought up by developers active in their online communities. While we extended previous research [8], it still leaves out significant areas of research such as conducting interviews, surveys, or analysis of other forums to gain insights.

There are limitations to our NLP analysis. The AdaBoost question classifier does not consider relations between tokens, parts of speech utilized in Reddit posts, or other potential features which leads to improper classification of rhetorical questions. For example, a post with: "Heard about a cool job posting? Let people know!" was misclassified. The sentiment analysis is also generic since the corpus used to generate the polarity scores, nltk.sentiment.vader, is specifically designed to be used with social media posts. While Reddit is a social media platform, a more directed corpus around software development most likely provides more accurate results. In the future, we will explore approaches proposed in [22, 35] to conduct a more detailed sentiment analysis. Lastly, our data collection for privacy-related questions is based on keywords search which may result in missing a large number of privacy content. We propose to use Natural Language Inference (NLI) approaches to extract a larger pool of data, similar to [36].

## 6. Conclusion and Future Work

In this paper, we looked into the developers' concerns and questions on Reddit forums to gain a better understanding of their attitudes toward privacy and the challenges they face when developing applications. We examined 437,317 threads from the subreddits r/webdev, r/androiddev, and r/iOSProgramming to determine the most frequently discussed topics as well as how the sentiments around these topics have evolved in response to GDPR and CCPA. Through a combination of word frequency, topic clustering, and question classifying, we observed that a large number of questions are related to requesting permissions from users, such as camera and location, or complying with the various regulations, particularly the GDPR. Additionally, we explored the emotions around dealing with privacy and found that there is a general neutral-to-positive sentiment around it across all types of development reviewed.

In the future, we plan to extend our effort to other developers' forums and improve our classification task. We will also leverage NLI to extract more privacy-related concepts.